\documentclass[12pt,preprint]{aastex}
\usepackage{natbib,color,rotating}
\newcommand{\myemail}{guoyang@nju.edu.cn}

\begin{document}
\title{Modeling Magnetic Field Structure of a Solar Active Region Corona using Nonlinear Force-Free Fields in Spherical Geometry}
\author{Y. Guo$^{1,2}$, M. D. Ding$^{1,2}$, Y. Liu$^{3}$, X. D. Sun$^{3}$, M. L. DeRosa$^{4}$, T. Wiegelmann$^{5}$}

\affil{$^1$ School of Astronomy and Space Science, Nanjing University, Nanjing
210093, China} \email{\myemail}

\affil{$^2$ Key Laboratory of Modern Astronomy and Astrophysics (Nanjing University), Ministry of Education, Nanjing 210093, China}

\affil{$^3$ W. W. Hansen Experimental Physics Laboratory, Stanford University, Stanford, CA 94305, USA}

\affil{$^4$ Lockheed Martin Advanced Technology Center, 3251 Hanover St., Palo Alto, CA 94304, USA}

\affil{$^5$ Max-Planck-Institut f\"ur Sonnensystemforschung, Max-Planck-Strasse 2, 37191 Katlenburg-Lindau, Germany}

\begin{abstract}

We test a nonlinear force-free field (NLFFF) optimization code in spherical geometry
using an analytical solution from Low and Lou. Several tests are run, ranging from idealized
cases where exact vector field data are provided on all boundaries, to cases where noisy
vector data are provided on only the lower boundary (approximating the solar problem). 
Analytical tests also show that the NLFFF code in the spherical geometry performs better 
than that in the Cartesian one when the field of view of the bottom boundary is large, say, 
$20^\circ \times 20^\circ$. Additionally, We apply the NLFFF model to an active 
region observed by the Helioseismic and Magnetic Imager (HMI) on board the \textit{Solar 
Dynamics Observatory} (\textit{SDO}) both before and after an M8.7 flare. For each 
observation time, we initialize the models using potential field source surface (PFSS) extrapolations
based on either a synoptic chart or a flux-dispersal model, and compare the resulting NLFFF
models. The results show that NLFFF extrapolations using the flux-dispersal model as the boundary condition
have slightly lower, therefore better, force-free and divergence-free metrics, and contain larger 
free magnetic energy. By comparing the extrapolated magnetic field lines with the extreme
ultraviolet (EUV) observations by the Atmospheric Imaging Assembly (AIA) on board \textit{SDO},
we find that the NLFFF performs better than the PFSS not only for the core field of the flare 
productive region, but also for large EUV loops higher than 50 Mm. 

\end{abstract}


\section{Introduction}

Magnetic field plays an important role in structuring the thermodynamic and hydrodynamic
behaviors of the plasma in the solar atmosphere. Three dimensional magnetic fields, especially
in the higher solar atmosphere, provide crucial information toward understanding various solar 
activities, such as filament eruptions, flares, and coronal mass ejections (CMEs). However, there 
are some difficulties in measuring the coronal magnetic field. The emissions and field strength 
are much weaker in the corona than that in the photosphere and chromosphere, while the
temperature is much higher producing a wide line width that covers the splitting by the Zeeman 
effect. Additionally, the observed intensities are integrations along the line-of-sight direction
due to the optically thin nature of the coronal plasma, and thus bright photospheric and chromospheric 
features obscure the emission originating in the corona. Although measurement of magnetic fields 
in the chromosphere and the corona has considerably improved in recent decades \citep{1998Judge, 
2000LinH, 2008Liu}, further developments are needed before accurate data are routinely available. 
Since photospheric magnetic fields are widely observed by various ground-based and space-borne 
instruments, and since the magnetic fields are in a low $\beta$ (the ratio between the gas pressure 
and the magnetic pressure) environment in the upper chromosphere, the transition region, and the lower
corona \citep{2001Gary}, a force-free field extrapolation technique provides an alternative 
way to infer the magnetic fields in the upper solar atmosphere.

The force-free magnetic field is described by the equations as follows:
\begin{eqnarray}
\nabla \times \mathbf{B} & = & \alpha \mathbf{B}~, \label{eqn1}\\
\nabla \cdot\mathbf{B}   & = & 0~, \label{eqn2}
\end{eqnarray}
where $\alpha$ is the torsion function that represents the proportionality between the
electric current density and the magnetic field. Combining equations (\ref{eqn1}) and 
(\ref{eqn2}), one obtains $\mathbf{B}\cdot\nabla \alpha = 0$, which indicates that $\alpha$
is always constant along a given field line. Setting $\alpha=0$ yields the current-free
potential field. For constant $\alpha$ over space, equations (\ref{eqn1}) and (\ref{eqn2}) describe the linear
force-free field. In the general nonlinear force-free field (NLFFF) that is closer to 
realistic coronal magnetic fields, $\alpha$ varies from one field line to another. Solutions
to the fully nonlinear problem can be calculated using numerical algorithms 
such as the Grad-Rubin, upward integration, MHD relaxation (magnetofrictional), optimization, 
and boundary element (Green's function like) methods (\citealt{2008Wiegelmann}, and references 
therein). A limited class of axisymmetric semi-analytical NLFFFs also exist 
\citep{1990Low}, and these are usually adopted as the standard test model of the above mentioned 
numerical algorithms, as has been done in \cite{2006Schrijver}. Many authors have shown that 
the force-free function $\alpha$ appearing in Equation~(\ref{eqn1}) changes in active regions 
\citep[e.g.,][]{2002Regnier,2005Schrijver}, thus the NLFFF model is necessary to construct 
the magnetic configuration for active regions, especially those with eruptive activities. 

NLFFF models have been adopted to study various magnetic field structures and properties 
in the solar atmosphere. For instance, \citet{2009Canou} found a twisted flux rope before
eruptive activities and \citet{2010Canou} obtained a flux rope as the magnetic structure
of a filament using the Grad-Rubin method. With the MHD relaxation method, \citet{2008Bobra}
and \citet{2011Su} constructed non-potential magnetic field models for solar active regions and 
a flare/CME event, respectively. The MHD relaxation method is implemented in spherical geometry.
A flux rope is usually present in this model since the flux rope is inserted into a potential
field beforehand. Using the optimization method, \citet{2010Guob,2010Guoa} found that a 
flux rope and magnetic arcades coexisted in a solar filament and they further computed 
the twist number of the flux rope and studied the eruption mechanism. \citet{2009Jing,2012Jing}
studied the evolution of free magnetic energy and relative magnetic helicity in activity 
productive active regions with the same optimization method. It is worth mentioning
some extrapolation methods without the force-free field assumption, for example, the non-force-free 
extrapolation of coronal magnetic field based on the principle of minimum dissipation rate 
(e.g., \citealt{2008Hu,2010Hu}) and the magnetohydrostatic modeling of the corona 
(e.g., \citealt{2003Wiegelmann,2007Wiegelmannb,2008Ruan}). Those methods should be more applicable
when the force-free state is no longer valid.

Most of the NLFFF procedures are implemented in the Cartesian coordinates. These procedures 
are not well suited for larger domains, since the spherical nature of the solar 
surface cannot be neglected when the field of view is too large. However, in a critical test
of NLFFF codes, \citet{2009DeRosa} concluded that the field of view should be as large as possible 
to include more information of magnetic field line connections. In addition to
the above reason, when we apply the optimization method for the NLFFF extrapolation to any 
realistically observed active region, an apriori requirement is that the magnetic field on 
the boundary is isolated \citep{2006Wiegelmann}. However, this requirement may not be satisfied 
in a realistic observation due to the limitation of the field of view. Therefore, it appears
prudent to implement a NLFFF procedure in spherical geometry for use when large-scale boundary
data are available, such as from the Helioseismic and Magnetic Imager (HMI; \citealt{2012Scherrer,
2012Schou}) on board the \textit{Solar Dynamic Obseratory} (\textit{SDO}). Furthermore, studies of 
magnetic field structures of large scale solar phenomena, such as trans-equatorial magnetic loops, 
large filaments, and CMEs, are also likely to benefit from NLFFF procedures in the spherical geometry. 

A NLFFF procedure in the spherical geometry based on the optimization method has been 
implemented by \citet{2007Wiegelmann}, which has been further developed by 
\citet{2009Tadesse} and \citet{2011Tadesse}. Another NLFFF procedure in the spherical 
geometry based on the same method has been implemented by J. McTiernan in the FORTRAN 90
language and released in the ``nlfff'' package available through Solar Software (SSW). 
Although \citet{2009Tadesse} have already tested the NLFFF code implemented by 
\citet{2007Wiegelmann} with the Low and Lou analytical solution, we test again J. McTiernan's 
version of the NLFFF code in this paper carefully. It is not only meaningful for the code 
itself before we apply it to real observations, but also an independent validation for the 
NLFFF method in spherical geometry.

We then apply the NLFFF procedure to a group of active regions observed on 2012 January 23 by 
\textit{SDO}/HMI. An M8.7 class flare occurred in NOAA 11402, one among the active region 
group. With this study, we first compare the extrapolated (both potential and NLFFF) magnetic 
loops with extreme ultraviolet (EUV) observations by the Atmospheric Imaging Assembly 
(AIA; \citealt{2012Lemen}) on board \textit{SDO}. This comparison indicates whether the 
NLFFF model reconstructs the magnetic configuration better than the potential field model. 
Additionally, this comparison can be used to evaluate how well the model field lines approximate
the observed coronal loops, thereby enabling a determination of whether the model is consistent
with observations. Furthermore, the spherical modeling technique enables the study of 
the non-potentiality of large magnetic loops and the magnetic surroundings of an active 
region, especially in a place where many active regions interact with each other. Finally, 
such a study can show how a flare, together with the associated filament eruption and CME,
changes the magnetic configuration permanently both on the photosphere and in the corona.

This paper is organized as follows. We first describe the procedure implementation in
Section~\ref{sec:optimization}. The testing of the procedure using the Low and Lou analytical 
solution, and comparisons between the NLFFF extrapolations in the Cartesian and spherical 
coordinates are presented in Section~\ref{sec:testing}. Then, we apply the NLFFF extrapolation 
in spherical geometry to \textit{SDO}/HMI observations and compare the NLFFF models with EUV
loops observed by \textit{SDO}/AIA in Section~\ref{sec:observation}. A summary and discussions 
are finally presented in Section~\ref{sec:conclusion}.

\section{Optimization Procedure in Spherical Geometry} \label{sec:optimization}

\citet{2000Wheatland} proposed an optimization method to reconstruct the NLFFF by minimizing
an objective functional that combines Lorentz forces and the divergence of the magnetic field.
If the functional is minimized to zero, equations (\ref{eqn1}) and (\ref{eqn2}) are satisfied 
simultaneously. The optimization procedure in the spherical geometry has been implemented by
\citet{2007Wiegelmann} and \citet{2009Tadesse}, where the objective functional is defined as
\begin{equation}
L_w = \int_{V} \omega (r,\theta,\phi)~\Bigl[B^{-2} \mid ( \nabla \times \mathbf{B} ) \times 
\mathbf{B}\mid^2 +  \mid \nabla \cdot \mathbf{B} \mid^2 \Bigr] r^2 \sin\theta \mathrm{d}r
\mathrm{d}\theta \mathrm{d}\phi ~,\label{eqn:obj}
\end{equation}
where $w(r,\theta,\phi)$ is a weighting function and $V$ is a wedge-shaped volume for numerical
computation. The weighting function is adopted to minimize the effects of the lateral and top
boundaries, where the data are not available in practical observations. The wedge-shaped volume 
is defined by six boundaries with $r \in [r_\mathrm{min},
r_\mathrm{max}]$, $\theta \in [\theta_\mathrm{min},\theta_\mathrm{max}]$, and $\phi \in 
[\phi_\mathrm{min},\phi_\mathrm{max}]$. The whole computational box is resolved by the grids 
of $n_r \times n_\theta \times n_\phi$.

For test cases with ideal boundary conditions, the vector field on all six boundaries can be 
specified by analytical solutions, which requires equal weights for all grid points in $V$ with 
$\omega (r,\theta,\phi) = 1$ everywhere. However, in more realistic cases, the vector field is
known only on the lower boundary. In these instances the boundary conditions on the remaining four
sides and top boundaries are typically specified by potential fields based on extrapolations of
the photospheric line-of-sight field. Because the potential field on the five boundaries 
does not match the NLFFF assumption exactly, the objective functional $L_w$ cannot decrease to 
zero. For these non-ideal cases we introduce a buffer region for 
the lateral and top boundaries, where the weighting function decreases from 1 to 0 with a cosine 
profile, from the inner volume $V'$ to the edges, in the buffer region with $n_\mathrm{buf}$ grid 
points. Thus, the inner physical domain $V'$ with weights of 1 spans $(n_r-n_\mathrm{buf}) 
\times (n_\theta-2 n_\mathrm{buf}) \times (n_\phi-2 n_\mathrm{buf})$ grid cells. The weighting 
function above enables departures of the field from the force-free and solenoidal state in the
buffer region, which allows the magnetic field in the inner physical domain $V'$ to evolve to a 
more force-free state than the code without a weighting function does. This is because
that those field lines with one of their footpoints on the lateral and top boundaries are half 
free, i.e., being allowed to change during the optimization procedure. The final state of the 
force-free field depends on the initial condition and the iteration history of the optimization 
procedure.

In this paper, we experiment with the optimization procedure in the Solar Software 
(SSW), which is implemented both in the Cartesian and spherical geometries and includes the weighting 
function, to reconstruct large scale coronal magnetic fields. This version of the optimization 
procedure that is fulfilled in the Cartesian coordinate system without weighting function has 
been tested in \citet{2006Schrijver} and \citet{2008Metcalf}. The author of the code is J. McTiernan. 
In the version that we adopt here, the procedure includes the weighting function as an input file.

\section{Testing the Optimization Procedure with the Low and Lou Solution} \label{sec:testing}

In this section, we test the optimization procedure as described in Section~\ref{sec:optimization}
using the \citet{1990Low} solution.
The subsections are arranged as follows. We first describe the settings of the Low and Lou
solution that is used as the standard test model in Section~\ref{sec:lowlou}. Secondly, the 
potential field source surface (PFSS) models that are used as initial conditions for the optimization 
method are discussed in Section~\ref{sec:pfss}. Thirdly, our testing results both for the ideal and noisy 
boundary conditions are presented in Section~\ref{sec:test}. Finally, we compare the extrapolation 
results derived by the NLFFF code in the Cartesian coordinates and spherical coordinates, respectively, 
in Section~\ref{sec:comparison}.

\subsection{Low and Lou Solution} \label{sec:lowlou}
The Low and Lou solutions describe a set of NLFFFs in axially symmetric geometries associated 
with point sources placed at the origin. Let us define the heliocentric coordinate system that is 
denoted by $O$-$xyz$ (also referred as the physical coordinate system), where $O$ is the center of the Sun and $xyz$ 
are the Cartesian components. If the origin of the local coordinate system, $O'$-$XYZ$, for 
the Low \& Lou solution is placed at ($-l_x$,0,$-l_z$) and the local $Y$-axis is parallel to 
the physical $y$-axis, the relation between the local coordinates and the physical ones is 
given by
\begin{equation}
\begin{array}{l}
X = (x+l_x) \cos \Phi - (z+l_z) \sin \Phi ~, \\
Y = y \ , \\
Z = (x+l_x) \sin \Phi + (z+l_z) \cos \Phi ~, \label{eqn4}
\end{array}
\end{equation}
where $\Phi$ is the angle counterclockwise measured from $z$-axis to $Z$-axis if viewed against 
$y$-axis. Although the transformation is modified a little as shown in Equation~\ref{eqn4} from
Equation~(13) of \citet{1990Low}, the coordinate system is identical to that depicted in Figure 2 of \citet{1990Low}.

\begin{figure}
\includegraphics[width=1.0\textwidth]{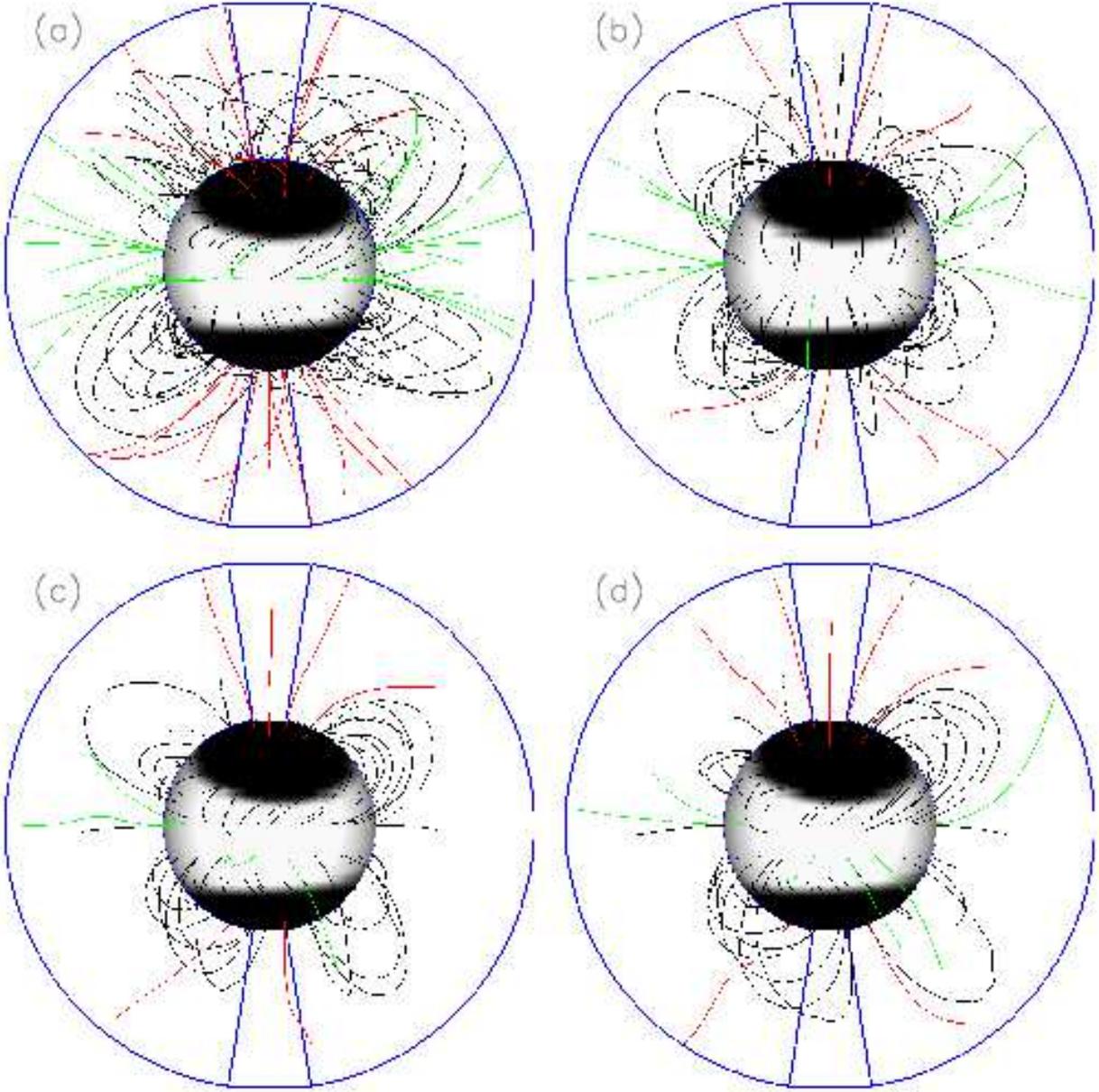}
\caption{(a) Low and Lou nonlinear force-free field solution. Red, green, and black
lines represent open field lines rooted on negative polarities, open field lines
rooted on positive polarities, and closed field lines, respectively. The blue line 
indicates a wedge shaped domain surrounded by $r \in [1.0 R_\sun,2.5 R_\sun]$, $\theta \in 
[9^\circ,171^\circ]$, and $\phi \in [90^\circ,270^\circ]$. All the metrics in 
Table~\ref{tbl:lowlou} are computed in this domain. (b) Potential field source surface model 
computed with the boundary of the Low and Lou solution. (c) Nonlinear force-free field
computed with the noisy boundary 1 (see texts for details). (d) Nonlinear force-free 
field computed with the preprocessed noisy boundary 2.} \label{fig:lowlou}
\end{figure}

We adopt the NLFFF solution with the eigenfunction of $P_{1,1}$ in the local coordinate 
system. $P_{1,1}$ is a solution of a nonlinear ordinary differential equation, corresponding
to parameters $n=1$ and $m=1$ of Equation (5) of \citet{1990Low}.
The final Low \& Lou solution in the physical volume with $r \in [1.0 R_\sun,2.5 R_\sun]$,
$\theta \in [0.0^\circ,180.0^\circ]$, and $\phi \in [0.0^\circ,360.0^\circ]$ is calculated with the 
parameters $l_x = 0.25 R_\sun$, $l_z = 0$, and $\Phi = -\pi /10$, and resolved by $30 \times 60 
\times 120$ grid points. All the three components $B_r$, $B_\theta$, and $B_\phi$ have been multiplied 
with a normalization constant to ensure that the maximum of $|B_r|$ equals 500 G. Some sample field 
lines of the Low \& Lou solution are shown in Figure~\ref{fig:lowlou}(a), where the central view 
point is located at $[\theta, \phi] = [90^\circ,180^\circ]$, i.e., an observer is observing along
the $x$-axis.

\subsection{Potential Field Source Surface Model} \label{sec:pfss}
A potential (current-free) field obeys the equation $\nabla \times \mathbf{B} = 0$, so that the
magnetic field can be expressed as the gradient of a scalar potential, i.e. $\mathbf{B} = - 
\nabla \Psi$. Since $\nabla \cdot \mathbf{B} = 0$, the scalar potential obeys the Laplace equation
$\nabla^2 \Psi = 0$. \citet{1969Schatten} introduced a spherical source surface at a distance $R_\mathrm{s}$
from the Sun's center, where the field lines are radial, to simulate the effect of the solar
wind on the magnetic field. Therefore, the field vectors are purely radial at $R_\mathrm{s}$, which indicates 
that $\Psi$ is a constant on the source surface. This constant can be selected as 0. Together with the 
photospheric boundary condition, which is traditionally provided by a map of the radial magnetic field,
the Laplace equation can be solved in the spherical coordinate system ($r, \theta, \phi$), where $\theta$ 
stands for colatitude. The solution in the domain $R_\sun < r < R_\mathrm{s}$ is
\begin{equation}
\begin{array}{l}
\Psi(r,\theta,\phi)=\sum\limits_{l=0}^{\infty}\sum\limits_{m=-l}^{l}[A_l^m r^l + B_l^m r^{-(l+1)}]Y_l^m(\theta,\phi), \label{eqn5}
\end{array}
\end{equation}
where the coefficients $A_l^m$ and $B_l^m$ are determined by the boundary conditions at $R_\sun$ and $R_\mathrm{s}$.
The spherical harmonic functions $Y_l^m(\theta,\phi)=C_l^m P_l^m(\cos \theta)e^{im\phi}$, where $C_l^m$
are normalization constants based on $l$ and $m$, and $P_l^m(\cos \theta)$ are associated Legendre functions.
Please refer to the appendix in \citet{2003Schrijver} for a detailed deduction of the coefficients and
the final magnetic field. Further information about spherical harmonic expansion of the Laplace equation
can also be found in \citet{1969Altschuler} and \citet{1977Altschuler}. Using the radial magnetic
field observed at $R_\sun$ and the source surface assumption at $R_\mathrm{s}$ as the boundary condition, together 
with the spherical harmonic expansion in the domain $R_\sun < r < R_\mathrm{s}$, we finally obtain the PFSS model.
Throughout this paper, the harmonic coefficients of the PFSS model are computed by the ``pfss'' package
available in SSW.

For the purpose of initializing the test runs based on the Low and Lou solution, we construct PFSS 
fields using the radial component of the magnetic field from the Low and Lou solution at the lower 
boundary. These fields are shown in Figure~\ref{fig:lowlou}(b). The principal order of the spherical 
harmonic series, $l_\mathrm{max}$, is set to 7. The solutions are smooth enough that higher $l$
values are not needed to resolve the functional form of the Low and Lou solutions at these radii.
We compute the correlation coefficient between the computed radial field ($b_r$) and the 
analytical radial field ($B_r$) and the normalized error (defined as $\sum |b_r-B_r| / 
\sum |B_r|$) on the bottom boundary as a measurement of the agreement between 
them. The correlation coefficient and normalized error of $b_r$ and $B_r$ are $1.00$ and $0.05$,
respectively, which indicates that the radial components of the analytical magnetic field are well 
reconstructed by the PFSS model.

\subsection{Test Results with Ideal and Noisy Boundary Conditions} \label{sec:test}

In this section, we will perform numerical experiments with the optimization procedure written 
by J. McTiernan in spherical geometry, with the aim of recovering the Low and Lou analytical 
solution described in Section~\ref{sec:lowlou} using different kinds of boundary conditions. 
Each experiment is initialized with potential fields calculated using 
the PFSS model described in Section~\ref{sec:pfss}. The NLFFF extrapolations are performed 
in a wedged-shaped domain bounded by $r \in [1.0 R_\sun,2.5 R_\sun]$, $\theta \in [9.0^\circ,171.0^\circ]$, 
and $\phi \in [90.0^\circ,270.0^\circ]$. This domain is resolved by $30 \times 54 \times 61$ grid points.
For the ideal boundary test cases, we arrange the following three different boundary conditions:
\begin{description}
  \item[Case 1:] The analytical vector magnetic fields from the $P_{1,1}$ solution of \citet{1990Low} 
are specified on all the six surfaces for the boundary condition.
  \item[Case 2:] The Low and Lou fields are specified only on the bottom layer, and the 
others are specified by the PFSS. No weighting function is used for lowering the boundary effect.
  \item[Case 3:] With similar boundary conditions used in Case 2, but a buffer zone with 6 grid points 
toward the top and lateral boundaries is adopted, where the weighting function decreases from 1 to 0 
with a cosine profile. The computation domain is the same size as Case 2, but with the buffer inside 
the Case 2 boundary.
\end{description}

In order to measure the convergence degree of the computed magnetic field, we define
the integral of the Lorentz force and the magnetic field divergence divided by the 
volume as follows:
\begin{equation}
L_\mathrm{f} = \frac{1}{V} \int_{V} \omega (r,\theta,\phi)~B^{-2} \mid ( \nabla \times 
\mathbf{B} ) \times \mathbf{B}\mid^2 r^2 \sin\theta \mathrm{d}r \mathrm{d}\theta \mathrm{d}\phi~, \label{eqn:forc}
\end{equation}
\begin{equation}
L_\mathrm{d} = \frac{1}{V} \int_{V} \omega (r,\theta,\phi) \mid \nabla \cdot \mathbf{B} \mid^2 
r^2 \sin\theta \mathrm{d}r \mathrm{d}\theta \mathrm{d}\phi ~. \label{eqn:dive}
\end{equation}
The magnetic field and the length units are Gauss and Mm, respectively. The two integrals 
should be 0 for perfect force-free and divergence-free field. The objective functional 
defined in Equation~(\ref{eqn:obj}) divided by volume $V$ is the summation of $L_\mathrm{f}$ 
and $L_\mathrm{d}$, i.e., $L_w = V (L_\mathrm{f} + L_\mathrm{d})$.

We use the figures of merit defined in \citet{2006Schrijver} and \citet{2008Metcalf} 
to quantify the degree of agreement between the analytical magnetic field $\mathbf{B}$ 
and the extrapolated $\mathbf{b}$. The vector correlation metric is defined as
\begin{equation}
C_\mathrm{vec} = \frac{\sum_i \mathbf{B}_i \cdot
\mathbf{b}_i}{\left(\sum_i B_i^2 \sum_i b_i^2\right)^{1/2}} ,
\label{eqn:vect}
\end{equation}
the Cauchy-Schwartz metric:
\begin{equation}
C_\mathrm{CS} = \frac{1}{M} \sum_i \frac{\mathbf{B}_i \cdot
\mathbf{b}_i}{B_i b_i} ,
\label{eqn:cauc}
\end{equation}
the normalized vector error metric:
\begin{equation}
E_\mathrm{N} = \frac{\sum_i |\mathbf{b}_i - \mathbf{B}_i|}{\sum_i B_i} , 
\label{eqn:norm}
\end{equation}
and the mean vector error metric:
\begin{equation}
E_\mathrm{M} = \frac{1}{M} \sum_i \frac{|\mathbf{b}_i -
\mathbf{B}_i|}{B_i} . \label{eqn:mean}
\end{equation}
Here, $B_i = |\mathbf{B}_i|$ and $b_i = |\mathbf{b}_i|$. The summations run over 
all the points $i$ in the volume of interest, and $M$ is the total number of points. 
If the two vector fields are identical, then $C_\mathrm{vec}=1$ and $C_\mathrm{CS}
=1$; $C_\mathrm{vec}=0$ and $C_\mathrm{CS}=0$ if $\mathbf{B}_i \perp \mathbf{b}_i$ 
at each point. Unlike the vector correlation and Cauchy-Schwarz metrics, $E_n=0$ 
and $E_m=0$ if the two vector fields are perfectly matched with each other. For a
consistent comparison with the other metrics, we list $1-E_\mathrm{n}$ and 
$1-E_\mathrm{m}$ in the following tables, so that all the metrics reach unity if 
the two vector fields are identical. Additionally, the magnetic energy in magnetic
field $\mathbf{b}$ normalized to that in the analytical magnetic field $\mathbf{B}$ 
is defined as
\begin{equation}
\epsilon = \frac{\sum_i b_i^2}{\sum_i B_i^2} ,
\label{eqn:ener}
\end{equation}
where $\epsilon = 1$ when the two fields are identical.

All the aforementioned metrics for the Low and Lou solution, the PFSS model with
$l_\mathrm{max}=7$, and the three test cases with ideal boundary conditions are listed 
in Table~\ref{tbl:lowlou}. The total energy contained in the corresponding magnetic 
field and that contained in the Low and Lou solution is also listed in the last column 
of Table~\ref{tbl:lowlou}. The PFSS model has lower $C_\mathrm{vec}$ and $C_\mathrm{CS}$ 
metrics, but larger $E_\mathrm{N}$ and $E_\mathrm{M}$ metrics compared with the Low and 
Lou solution. The PFSS magnetic energy is only $82 \%$ of the reference one. However, 
using the potential field as the initial condition, the NLFFF with all the six boundary 
vector magnetic fields are provided (Case 1) is recovered to the Low and Lou solution 
with very high accuracy as shown in Table~\ref{tbl:lowlou}. For a more practical case, 
in which only the bottom vector magnetic fields are known, while the others are specified by the 
PFSS model, and we do not use the weighting function (Case 2), the results
show that $L_\mathrm{f}$ and $L_\mathrm{d}$ are one order of magnitude larger than that
of the Low and Lou solution. However, the $C_\mathrm{vec}$, $C_\mathrm{CS}$, $E_\mathrm{N}$, and 
$E_\mathrm{M}$ metrics are better than that of the PFSS model. The magnetic energy in 
Case 2 is recovered to $92 \%$ of the energy in the Low and Lou solution. If a weighting 
function decreasing from 1 to 0 with a cosine profile is adopted in the buffer zone (Case 3),
$L_\mathrm{f}$, $L_\mathrm{d}$, and the magnetic energy have similar values as that in Case 1.
The $C_\mathrm{vec}$, $C_\mathrm{CS}$, $E_\mathrm{N}$, and $E_\mathrm{M}$ metrics are worse 
than that in Case 1 but better than that in Case 2. Therefore, a weighting function for those
boundaries where data are missing improves the NLFFF extrapolation.

\citet{1969Molodensky,1974Molodensky}, \citet{1989Aly}, and \citet{1989Sakurai} pointed out
that vector magnetic fields on a closed surface that fully encloses any force-free domain
have to satisfy the force-free and torque-free conditions. In practical observations, only
data on the bottom boundary are available. Therefore, the force-free and torque-free conditions
are required in a well isolated region in a force-free magnetic field. In an isolated region,
all field lines originating from the bottom boundary fall back on it again. The isolation 
condition requires that the flux is in balance, an area with strong magnetic fields 
(active region) is surrounded by weak fields (quiet Sun), and no flux interconnects to other 
active regions. However, the flux balance is a necessary 
condition for the isolation of the magnetic field, but it is not a sufficient one, which
implies that it is better to enlarge the field of view than to cut a smaller region in order
to fulfill the isolation condition. On the photosphere the force-free and torque-free conditions
are not well fulfilled in an isolated region since the plasma $\beta$ is close to unity 
\citep{2001Gary}. Thus, the forced photospheric boundaries need to be preprocessed. 
\citet{2006Wiegelmann} proposed a preprocessing routine to remove the net magnetic force and
net magnetic torque, and to smooth the magnetic field, while keeping the changes of the magnetic
field within the noise level. \citet{2009Tadesse} further developed the preprocessing routine
into the spherical geometry. Following the idea of \citet{2009Tadesse}, we have written a
version of the preprocessing code and applied it to the following analysis.

We add Gaussian-distributed random noise to the bottom boundary of the Low and
Lou solution to simulate realistic observations. For each component of the magnetic
field ($B_r$, $B_\theta$, or $B_\phi$) at position $i$, a noisy component $\delta B_\gamma 
= l_\gamma \cdot r_i$ ($\gamma = r, \theta$ or $\phi$) is added to the original magnetic field, 
where $l_\gamma$ is the level of noise for that component, and $r_i$ is the random number in 
the range of $-1$ to $1$ for position $i$. For all the grid points, $r_i$ has a Gaussian distribution. 
By choosing the noise level $l_\gamma$, we have two different test cases here:
\begin{description}
  \item[Noisy model 1:] $l_r = 0.1 \times \sqrt{\mathrm{max}(|B_r|)}$, $l_\theta = \sqrt{\mathrm{max}(|B_\theta|)}$,
and $l_\phi = \sqrt{\mathrm{max}(|B_\phi|)}$.
  \item[Noisy model 2:] $l_r = 0.2 \times \sqrt{\mathrm{max}(|B_r|)}$, $l_\theta = 2 \times \sqrt{\mathrm{max}(|B_\theta|)}$,
and $l_\phi = 2 \times \sqrt{\mathrm{max}(|B_\phi|)}$.
\end{description}
The two noisy models are similar to noisy model II of \citet{2006Wiegelmann},
i.e., adding additional noises independent of the magnetic field strength. We only choose
a different strategy to determine the noise level for each component of the magnetic field.
The difference between noisy model 1 and noisy model 2 is that the latter has a noise 
level that is twice as large as that of the former.

For each of the two noisy models, we apply the preprocessing procedure to remove the net magnetic
force and net magnetic torque, and to smooth the magnetic field. The preprocessed bottom boundaries are
submitted to the optimization procedure for NLFFF extrapolations. The other settings are the same as
the test case 3 with an ideal bottom boundary, i.e., the initial condition and the boundary condition
for the other five boundaries (other than the bottom one) are specified by the PFSS model. The 
extrapolation is also performed in the wedged shaped domain bounded by $r \in [1.0 R_\sun,2.5 R_\sun]$, 
$\theta \in [9.0^\circ,171.0^\circ]$, and $\phi \in [90.0^\circ,270.0^\circ]$, which is resolved by $30 
\times 54 \times 61$ grid points. For comparison, the two noisy boundaries without preprocessing 
are also submitted for additional NLFFF extrapolations. The metrics for the extrapolation results
are shown in the last four rows in Table~\ref{tbl:lowlou}. It shows that all the metrics for both noisy models 
1 and 2 are improved after preprocessing of the bottom boundary. Since the normalized vector error 
metric $E_\mathrm{N}$ and the mean vector error metric $E_\mathrm{M}$ are more sensitive to the 
difference of two sets of vector fields, we focus on the changes of these two metrics for the two
noisy models before and after the preprocessing. The changes of $E_\mathrm{N}$ and $E_\mathrm{M}$ show
that if the noises are larger in the bottom boundary, the preprocessing could improve the extrapolation
results more significantly.

\subsection{Comparison between Cartesian and Spherical Extrapolations} \label{sec:comparison}

\begin{figure}
\includegraphics[width=1.0\textwidth]{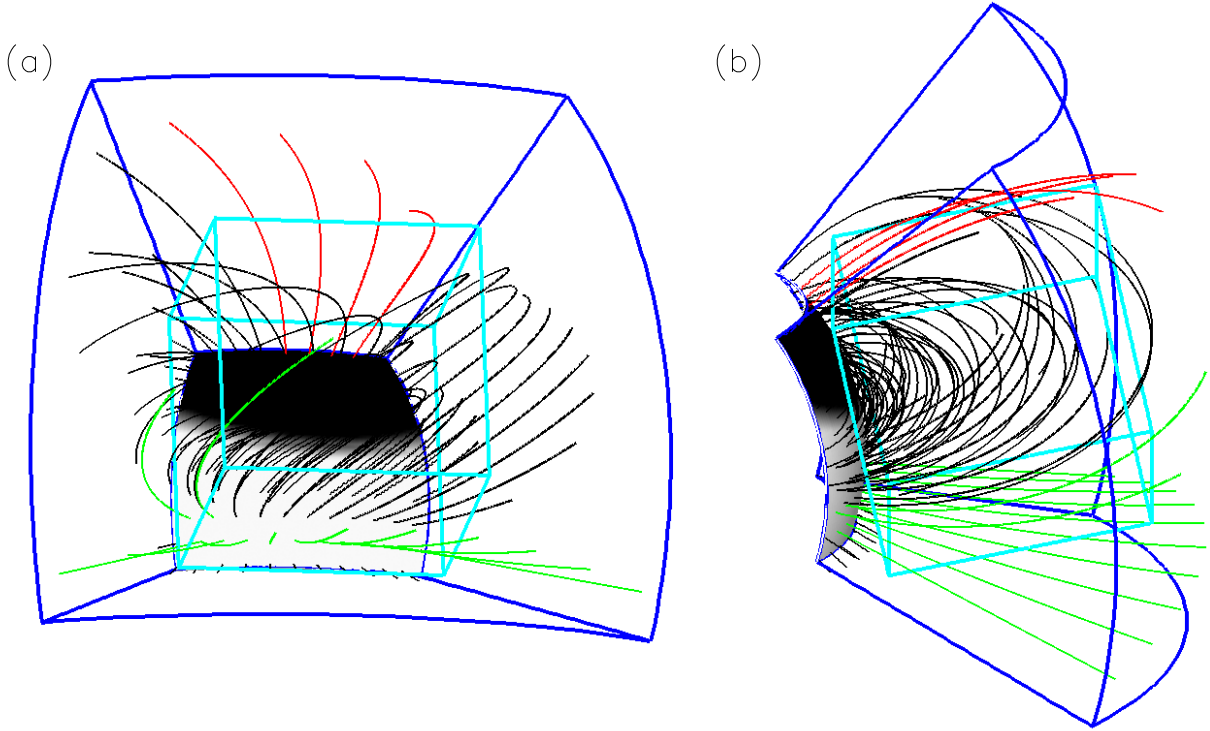}
\caption{A wedge shaped region cut from the Low and Lou solution. The center of the bottom
surface is located at $[\theta, \phi] = [78^\circ, 180^\circ]$. Blue lines mark the boundary
of the wadge shaped region. Cyan lines represent a cubic box that is tangent to the solar 
surface at $[r, \theta, \phi] = [1 R_\sun, 78^\circ, 180^\circ]$. Red, green, and black
lines represent open field lines rooted on negative polarities, open field lines
rooted on positive polarities, and closed field lines, respectively. (a) The view angle
is $[\theta, \phi] = [100^\circ, 170^\circ]$. (b) The view angle is $[\theta, \phi] = 
[70^\circ, 90^\circ]$.} \label{fig:cartesian}
\end{figure}

We cut a cubic box in the Low and Lou analytical magnetic field. The bottom surface of the 
box is tangent with the solar surface at $[r, \theta, \phi] = [1.0 R_\sun, 78.0^\circ, 180.0^\circ]$.
The length of each side, $l$, is determined by $l = 2 R_\sun \tan (\theta_l / 2)$, where $\theta_l$
is a free parameter indicating the angle of the box side viewed from the center of the Sun. 
For example, Figure~\ref{fig:cartesian} shows the case where $\theta_l = 60^\circ$. We prepare 
three cases with $\theta_l = 20^\circ$, $40^\circ$, and $60^\circ$. For each case, we do two 
NLFFF extrapolations in the spherical and Cartesian coordinates, respectively. The bottom boundary 
for the NLFFF extrapolation in the spherical geometry is the analytical magnetic field at $1 R_\sun$ 
and bounded by the longitude and latitude lines at the four extremities of the bottom surface of the 
cubic box. The bottom boundary for the Cartesian extrapolation is also the analytical magnetic 
field at $1 R_\sun$, but the geometry of the magnetic fields are projected to the tangent bottom 
surface of the cubic box, while the vector components are kept in the heliographic coordinates. 
The other five boundaries for both cases are given by the potential field.
Such a way to prescribe the boundary condition mimics real observations, since we only have real 
observations on the bottom boundary at present and people usually assume that the heliographic
components are equivalent to the Cartesian components when the field of view is small. 

We adopt a buffer zone with 6 grid points for the weighting function. The preprocessing procedure 
is not applied to the boundaries. The NLFFF extrapolations are performed in the cubic box
and the wedge shaped region for the Cartesian and spherical coordinate cases, respectively. The metrics for 
both the extrapolations in the Cartesian and spherical geometries are listed in Table~\ref{tbl:cartesian}.
The force-free measure, $L_\mathrm{f}$, and the divergence-free measure, $L_\mathrm{d}$,
are computed in the corresponding computation domain, i.e., the cubic box for the Cartesian
case and the wedged shaped region for the spherical one. 
All the other metrics are computed in the inner region (excluding the six-grid buffer region) of the 
corresponding cubic box. Note that even for the spherical extrapolations, only the magnetic field in
the inner region of the corresponding cubic box is extracted for computing the metrics. All the 
metrics in Table~\ref{tbl:cartesian} show that the NLFFF code in the spherical geometry performs 
better than that in the Cartesian one when the field of view of the bottom boundary is larger than 
$20^\circ \times 20^\circ$.

\section{Applications of the Spherical NLFFF to Observations} \label{sec:observation}

In order to apply the NLFFF model to observations, we need vector magnetic field measurements in
a relatively large field of view and the initial condition that is often from the PFSS model,
which is computed by the boundary condition from a global radial magnetic field. The
\textit{SDO}/HMI provides us such magnetic fields with its full disk, high resolution, and high 
cadence vector magnetograms.

\subsection{Instrument and Data Analysis}

\textit{SDO}/HMI observes the full Sun with a $4 \mathrm{K} \times 4 \mathrm{K}$ CCD, whose spatial
sampling is $0.5''$ per pixel. It obtains raw filtergrams at six different wavelengths and six
polarization states (i.e., $I \pm S$, where $S = Q, U$, and $V$) in the Fe {\sc \romannumeral 1} 
6173 \AA \ spectral line. It takes 135 s to acquire each set of the filtergrams for deriving 
the vector magnetic field. All the filtergrams are averaged over 12 minutes to increase the 
signal to noise ratio and to remove the p-mode signals. The four Stokes parameters, $I, Q, U$, 
and $V$, at the six wavelengths are computed from the raw data after necessary calibrations, such 
as flat field correction, dark frame subtraction, and so on.

The vector magnetic field, together with other thermodynamical parameters, are computed by the 
code of Very Fast Inversion of the Stokes Vector (VFISV; \citealt{2011Borrero}), which fits 
the synthesized Stokes profiles to the observed ones by a least square fitting method. VFISV 
adopts a line synthesized model that is based on the Milne-Eddington atmosphere, where all
the physical parameters are constant along the line of sight, except for the source function,
which depends linearly on the optical depth. VFISV has been optimized for the HMI data preprocessing;
for instance, the damping constant has been set to 0.5 and the filling factor to 1. Therefore,
the atmosphere is assumed to be uniformly filled with magnetic fields in each pixel. 

The transverse components of vector magnetic fields suffer from the so-called $180^\circ$ ambiguity. 
The $180^\circ$ ambiguity for the HMI data in this study has been resolved by an improved version 
of the minimum energy method \citep{1994Metcalf,2006Metcalf,2009Leka}. As described in 
\citet{2009Leka}, in weak-field areas, the minimization may not return a good solution due to
large noise. Therefore, in order to get a spatially smooth solution in weak-field areas, we 
divide the magnetic field into two regions, i.e., strong-field region and weak-field region,
which is defined to be where the field strength is below 200 G at the disk center, and 400 G on the 
limb. The values vary linearly with distance from the center to the limb. The ambiguity solution in the
strong-field area is derived by the annealing in the minimization method and released in the data
series ``hmi.B\_720s\_e15w1332\_CEA''. Magnetic field azimuths in the weak-field area are finally 
determined by the potential-field acute-angle method after the annealing. This is different from 
the original minimum energy method, which uses a neighboring-pixel acute-angle algorithm to revisit 
the weak-field area.

If a magnetic field vector is observed far from the solar disk center, the projection effect 
distorts the geometric shape of observed data, and the line-of-sight and the two transverse 
components ($B_\xi, B_\eta$, and $B_\zeta$) decomposed in the image plane coordinates deviate 
from the heliographic east--west, south--north, and vertical components ($B_x, B_y$, and $B_z$). 
Therefore, correction for the projection effect must include two aspects. One corrects the 
geometric shape of the observed data, which is often called remapping \citep{2002Calabretta,
2006Thompson}. The other transforms the magnetic field components in the image plane to those 
in the heliographic plane \citep{1990Gary}. The heliographic components $B_x, B_y$, and $B_z$ 
are equivalent to the spherical components $B_\phi, -B_\theta$, and $B_r$. 

\begin{figure}
\includegraphics[width=1.0\textwidth]{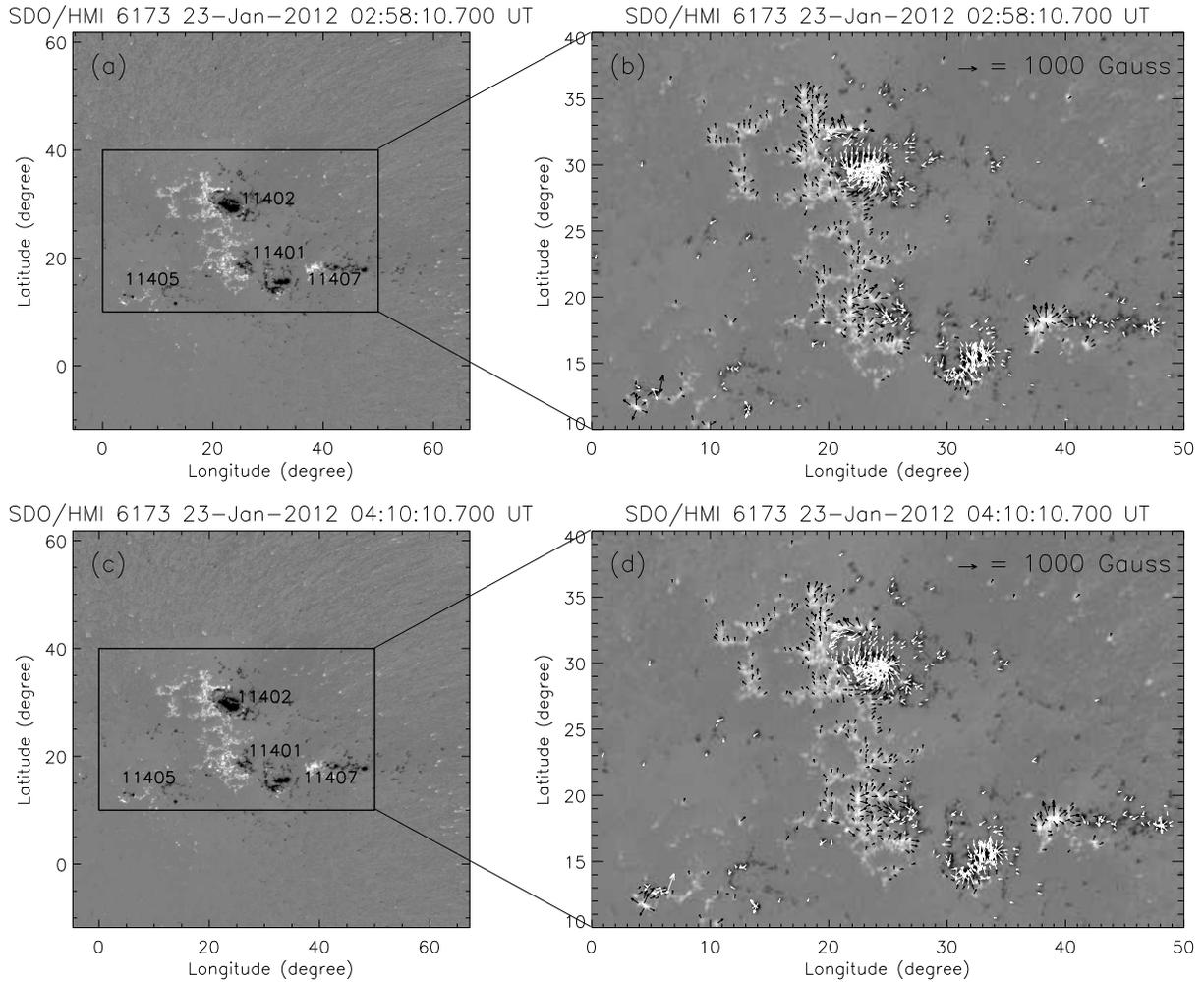}
\caption{Vector magnetic fields observed by \textit{SDO}/HMI. (a) The vertical component of the vector 
magnetic field observed at 02:58 UT on 2012 January 23. The black rectangle represents the field
of view for panel (b). Written in the rectangle are the NOAA numbers for the active regions. 
(b) The vector magnetic field observed at 02:58 UT in the field of view as shown in the rectangle
of panel (a). The arrows stand for the horizontal components.
(c) The vertical component of the vector magnetic field observed at 04:10 UT.
(d) The vector magnetic field observed at 04:10 UT.} \label{fig:vector}
\end{figure}

Figure~\ref{fig:vector} shows two sets of remapped and reprojected vector magnetic fields 
observed by HMI at 02:58 UT and 04:10 UT on 2012 January 23. The Joint Science Operations 
Center (JSOC) series name for the data is ``hmi.ME\_720s\_e15w1332''. More detail about the data
are described in http://jsoc.stanford.edu/jsocwiki/VectorMagneticField. An M8.7 class flare peaked at 
03:38 UT in active region NOAA 11402 on the same day. There are three additional active regions, 
NOAA 11401, 11405, and 11407 that interacted with NOAA 11402 in this area. The field of view 
covering all the active regions is about $71.7^\circ \times 71.5^\circ$. Due to the large 
field of view, the NLFFF in the spherical geometry is necessary for studying the magnetic field 
configuration of the flaring active region and the magnetic connections between these active regions.

\subsection{Boundary and Initial Conditions} \label{sec:bcic}

Because vector-field measurements are not available for the side and top boundaries of a localized
domain, one needs to make assumptions about these fields before performing a NLFFF extrapolation.
Here, we assume the lateral and upper boundaries of the computational domain are current-free,
and make use of the well-known PFSS model to calculate the boundary conditions. In order to 
include as much information of magnetic field connections as possible and conform 
to the requirement of the ``pfss'' package, we need a global radial magnetic field as the bottom 
boundary for PFSS computations.
But for realistic observations up to now, we could only get the full disk magnetogram, or the full 
disk vector magnetic field, in the best situation. It means that there are at least the following 
problems to be solved in order to construct a global radial magnetic field. 

First, we do not have magnetic field observations on the backside of the Sun. Therefore, the global 
magnetic field is usually constructed by the synoptic map, which is made by consecutive slices of
magnetograms when each of them passes the central meridian. The slices of magnetograms are remapped
on the heliographic plane, and are arranged side by side according to their Carrington coordinates. 
A synoptic map implicitly neglects the differential rotation of the solar surface, and only has a 
temporal resolution of the Carrington rotation period, i.e., approximately 27 days. In order to study 
the change of large-scale magnetic structures on the time scale much less than the Carrington rotation 
period, \citet{1999Zhao} constructed a so-called ``magnetic frame'' consisting of two components, i.e., 
a classic synoptic map inserted with a remapped magnetogram observed at a short time range.

Secondly, the magnetic field in the polar region of the Sun is not presently well observed. The poles are 
blocked periodically due to the tilt angle of about $7.25^\circ$ between the Sun's equatorial plane 
and the ecliptic plane. The radial fields in the polar region are approximately transverse to lines
of sight as observed by instruments near the ecliptic plane, which introduces high noises. 
However, the large-scale polar magnetic fields are essential to build the PFSS model; therefore,
they are important for the NLFFF modeling. \citet{2011Sun} proposed a new two-dimensional spatial/temporal 
interpolation and flux transport model to interpolate the desired data. When we use the observed 
data (synoptic map or synoptic frame) as the boundary condition for the NLFFF extrapolation in the 
following sections, we adopt the new interpolation method to fill the magnetic field in the polar 
regions.

Finally, a synoptic map or synoptic frame still lacks some basic physics of the magnetic field 
behavior on the solar surface, such as the differential rotation and the evolution of the magnetic
field during a Carrington rotation. \citet{2001Schrijver} and \citet{2003Schrijver} developed a
flux-dispersal model and a data-assimilation procedure to address the aforementioned problems. The 
flux-dispersal model advects the magnetic flux across the solar surface and incorporates all the
ingredients that are necessary for the magnetic flux evolution, such as the differential rotation,
meridional flow, convective dispersal, flux emergence, flux fragmentation, and flux cancellation.
\textit{SDO}/HMI or the Michelson Doppler Imager (MDI) line-of-sight magnetograms are assimilated 
into the model when the data are available.

We apply the PFSS model to two different kinds of photospheric radial magnetic field data as 
described above. One is the synoptic frame \citep{1999Zhao} and the other is an evolving flux 
dispersal model \citep{2001Schrijver, 2003Schrijver} that makes use of magnetograms assimilated 
into the model. These global magnetic maps are used to construct the PFSS models that are used 
for the initial fields, and for the boundary conditions on the sides and top for both the selected 
observation times. The steps to construct the global radial magnetic field for each observation 
time are detailed as follows.

First, the two vector magnetic fields as shown in Figure~\ref{fig:vector} are interpolated to 
$144 \times 144$ grid points uniformly distributed in latitude and longitude with a grid spacing 
of about $0.5^\circ$ per pixel. The vector magnetic fields are then preprocessed to remove the 
net magnetic force, torque, and noises using the preprocessing method in the spherical geometry 
as discussed in Section~\ref{sec:test}. 

\begin{figure}
\includegraphics[width=1.0\textwidth]{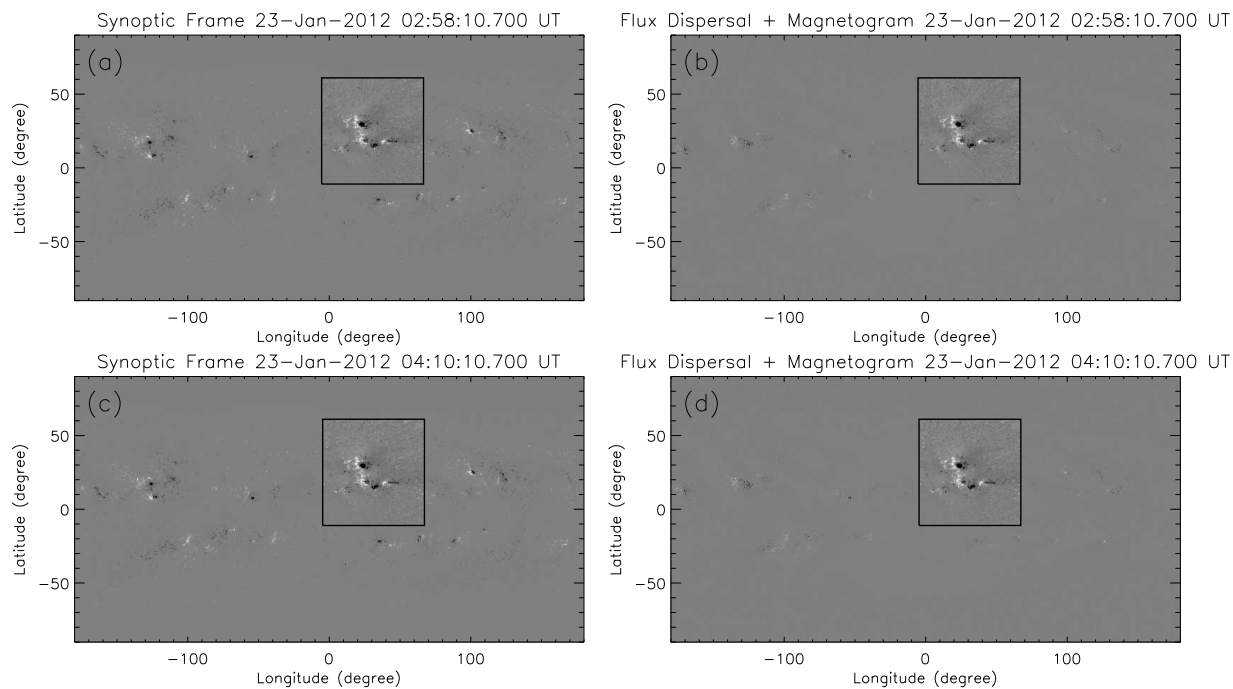}
\caption{(a) The synoptic frame that is a combination of the synoptic map for the 
Carrington Rotation 2119 (2012 January 9 to February 6) and the magnetogram observed by \textit{SDO}/HMI 
at 02:58 UT on 2012 January 23. The black rectangle indicates the field of view of the magnetogram. 
(b) The global vertical magnetic field that is a combination 
of the flux dispersal model at 00:04 UT on 2012 January 23 and the magnetogram observed by \textit{SDO}/HMI 
at 02:58 UT on 2012 January 23. (c) Similar to panel (a) but using the magnetogram at 04:10 UT.
(d) Similar to panel (b) but using the flux dispersal model at 06:04 UT and the magnetogram 
at 04:10 UT.} \label{fig:global}
\end{figure}

Secondly, the synoptic map for the Carrington Rotation 2119 (2012 January 9 to February 6) 
is interpolated onto a uniform $720 \times 360$ pixel grid in latitude and longitude. 
The synoptic map is produced by line-of-sight magnetic field 
observations, which have been converted to the radial component by assuming that the
magnetic field is radial. This synoptic map is used for both the observations at 02:58 UT and 
04:10 UT, as we assume that the magnetic field outside the group of active regions did not 
change too much during this period. The synoptic map is aligned with the vector magnetic 
fields by a correlation method. Then, the region where the \textit{SDO}/HMI data are available 
is replaced by the preprocessed radial magnetic field. These new maps, or synoptic frames, are 
shown in Figures~\ref{fig:global}(a) and \ref{fig:global}(c) for the observations at 02:58 UT 
and 04:10 UT, respectively.

As an alternative, we prepare another version of the global radial magnetic field using an evolving flux 
dispersal model. For the observation at 02:58 UT, the surface radial magnetic field of 
the flux dispersal model at 00:04 UT is interpolated to $720 \times 360$ grid points, 
and aligned with the radial component of the preprocessed \textit{SDO}/HMI vector 
magnetic field at 02:58 UT. Additionally, we combine 
the aligned flux dispersal field with the \textit{SDO}/HMI radial field to include more 
accurate information where the observed data are available. Compared to the magnetic 
frame, the flux dispersal model inserted with the observed data can be sampled every
6 hr and thus may provide a more consistent bottom boundary, since the flux dispersal 
model has considered the differential rotation, meridional flow and other necessary factors. 
For the observation at 04:10 UT, we use the flux dispersal model at 06:04 UT. 
Figures~\ref{fig:global}(b) and \ref{fig:global}(d) show the flux dispersal model
inserted with the \textit{SDO}/HMI magnetogram for the observations at 02:58 UT and 
04:10 UT, respectively.

\begin{figure}
\includegraphics[width=1.0\textwidth]{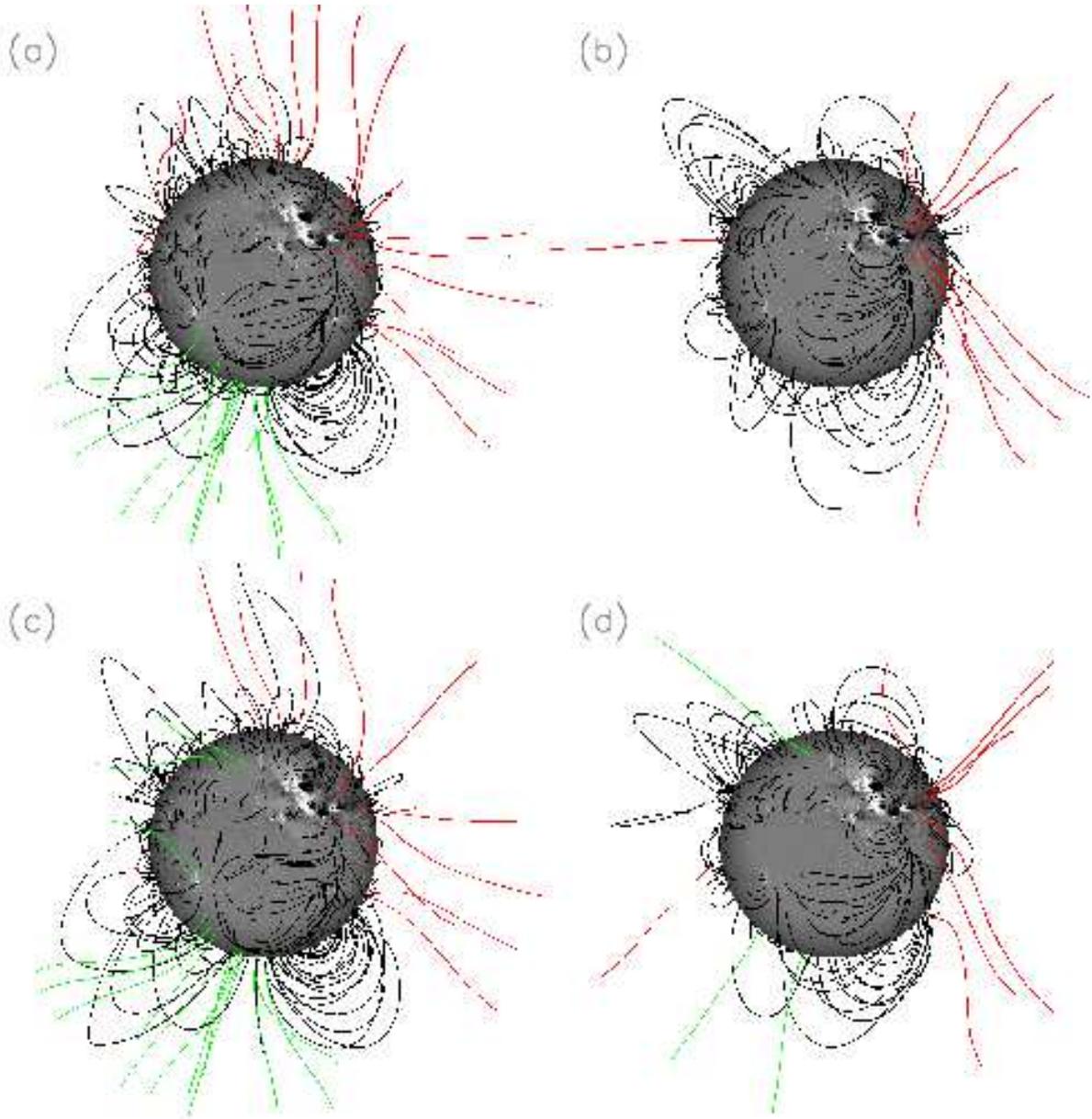}
\caption{The PFSS models computed with the four different boundaries as shown in Figure~\ref{fig:global}.
Panels (a)--(d) and those in Figure~\ref{fig:global} have a one-to-one correspondence. Red, green, and black
lines represent open field lines rooted on negative polarities, open field lines
rooted on positive polarities, and closed field lines, respectively.} \label{fig:pfss}
\end{figure}

The PFSS models based on the two different boundary conditions at the two different times 
are computed with $l_\mathrm{max}=360$. When the radius is large, a smaller $l_\mathrm{max}$ 
is used. The correlation coefficients between the radial fields of the PFSS models 
($b_r$) on the bottom boundary and the corresponding global radial fields ($B_r$) 
for all the four cases are between $0.93$ to $0.96$. The normalized errors of $b_r$ and
$B_r$ (defined as $\sum |b_r-B_r| / \sum |B_r|$) are between $0.27$ to $0.34$. Figure~\ref{fig:pfss} shows a 
sample set of field lines traced through the PFSS models, which will later serve both
as the side and top boundary conditions as well as the initial conditions for the NLFFF 
extrapolation in spherical geometry. Comparing the PFSS models at two observation 
times, we find that the PFSS models computed with the magnetic frame have evolved less 
than those computed with the flux-dispersal model with the inserted magnetogram. 
The reason is that the synoptic map used for the two observation times is the same.
Comparing the PFSS models with the two different boundaries at each time, we find that
the polar fields computed with the magnetic frames are stronger than that with the 
other boundary, and this is why the ``hairy Sun'' images in Figure~\ref{fig:pfss}
look different. The force-free, 
divergence-free measures, and the magnetic energy for the PFSS models are listed in 
Table~\ref{tbl:hmi}. They have the same value for the two different boundaries for 
each observation time.

\subsection{Results} \label{sec:result}

The NLFFF models for both the observation times are derived via the optimization method
with the boundary and initial conditions specified by the \textit{SDO}/HMI vector magnetic
fields and the PFSS models applied to both the synoptic frame and the flux-dispersal model. 
Table~\ref{tbl:hmi} lists the force-free, divergence-free measures, and the magnetic 
energy contained in these models. As discussed in Section~\ref{sec:bcic}, using
different boundaries for the PFSS model does not affect the force-free, divergence-free
measures, and magnetic energy for the derived PFSS model. However, as we use these
PFSS models for the NLFFF extrapolation, the flux-dispersal model gives better force-free,
divergence-free measures, and larger magnetic energy than that computed from the synoptic
frame for both the observation times. Since the potential field magnetic energy is the
same, the free magnetic energy derived from the flux-dispersal model at 02:58 UT ($2.0
\times 10^{32}$ ergs) is thus larger than that from the synoptic frame ($1.5 \times 10^{32}$ ergs).
Similarly, the magnetic energy at 04:10 UT from the flux-dispersal model ($1.2 \times 
10^{32}$ ergs) is also larger than that from the synoptic frame ($1.0 \times 10^{32}$ ergs).
The difference in free energies between the two times, which we assume corresponds to the
energy released by the M8.7 flare, is $0.8 \times 10^{32}$ ergs
or $0.5 \times 10^{32}$ ergs for the flux-dispersal model or the synoptic frame as the
boundary conditions, respectively. 

The errors listed in Table~\ref{tbl:hmi} are estimated by a pseudo Monte Carlo experiment. 
For each of the four cases in Table~\ref{tbl:hmi}, We add some Gaussian-distributed random noise
to the global radial magnetic field. In order to focus on the effects of the two global magnetic 
field models, i.e., the synoptic map and the flux-dispersal model, the random errors are not
added to the vector magnetic field. The error level (the standard deviation of the random noises) 
is selected to be 10 G \citep{2012Hoeksema}. We create 10 noisy global radial magnetic fields for 
each case as listed in Table~\ref{tbl:hmi}, combine them with the magnetogram observed by 
\textit{SDO}/HMI, and do the PFSS and the NLFFF extrapolations. The metrics listed in Table~\ref{tbl:hmi} 
are the mean values derived from the extrapolation results computed from the 10 noisy and 1 
original boundaries. The errors are estimated as the standard deviations. 

\begin{figure}
\includegraphics[width=1.0\textwidth]{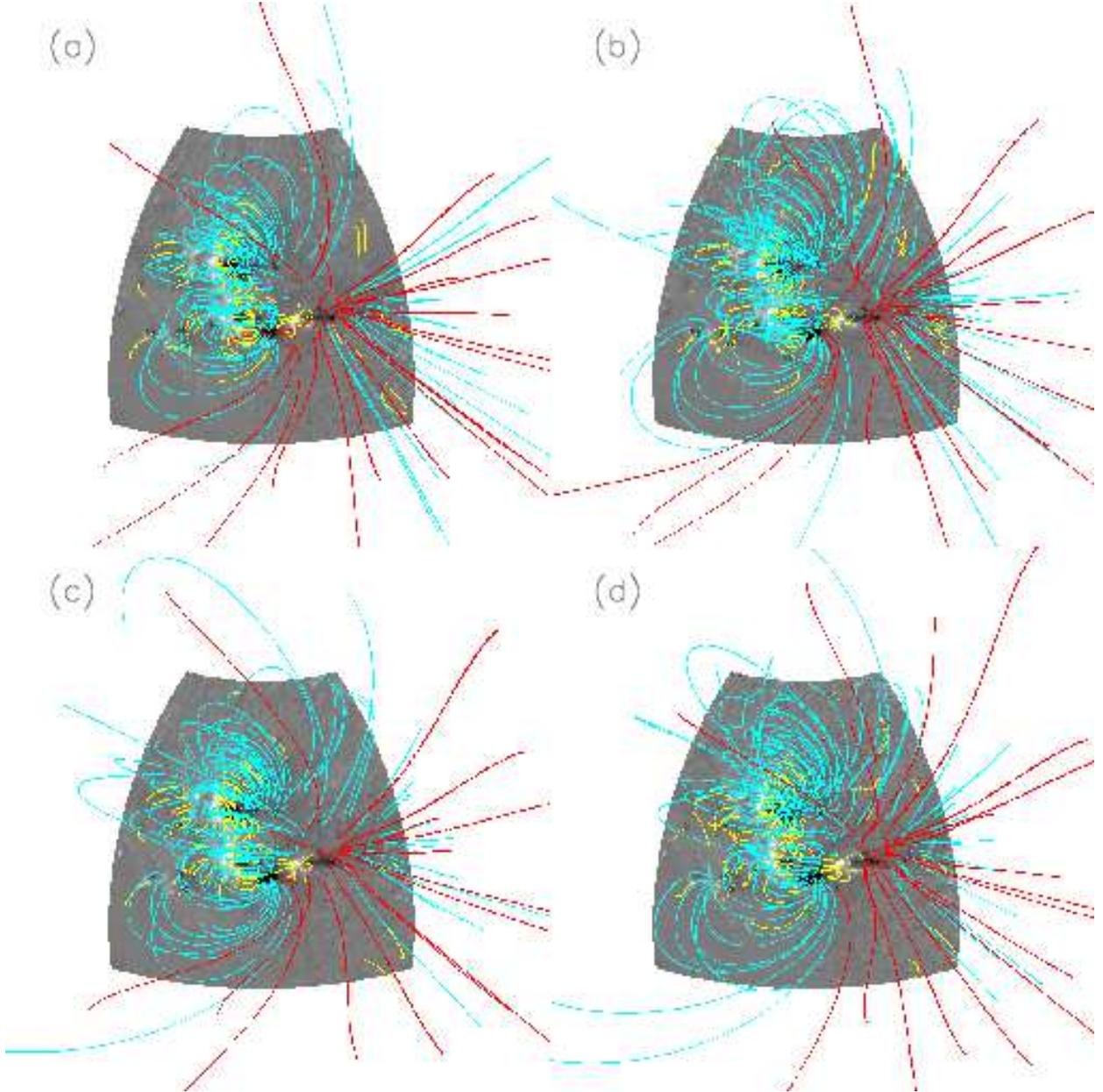}
\caption{(a) The PFSS model extracted from panel (b) of Figure~\ref{fig:pfss} in the domain of 
$r \in [1.0 R_\sun,2.5 R_\sun]$, $\theta \in [29.1^\circ,100.8^\circ]$, and $\phi \in [-5.0^\circ,66.5^\circ]$. 
The yellow lines indicate the magnetic field lines, whose maximum heights are less than $1.06 R_\sun$.
The cyan lines are for other closed magnetic field lines.
The red lines represent open magnetic field lines. 
(b) The NLFFF model computed by the vector magnetic field observed at 02:58 UT and the PFSS model
in panel~(a). 
(c) The PFSS model extracted from panel (d) of Figure~\ref{fig:pfss} in the domain of 
$r \in [1.0 R_\sun,2.5 R_\sun]$, $\theta \in [29.1^\circ,100.8^\circ]$, and $\phi \in [-4.5^\circ,67.0^\circ]$. 
(d) The NLFFF model computed by the vector magnetic field observed at 04:10 UT and the PFSS model in panel~(c). 
} \label{fig:nlfff}
\end{figure}

We plot the PFSS and NLFFF models in Figure~\ref{fig:nlfff} in the wedge shaped field of 
view for the two observation times. Note that only the NLFFF extrapolations from the 
flux-dispersal model are plotted, since this model yields smaller force-free 
and divergence-free metrics. In these figures, the viewer is located above the center
of the computational domain. For each observation time in Figure~\ref{fig:nlfff}, the 
field lines of PFSS and NLFFF models are integrated from the same footpoints. We find
that some field lines of the NLFFF at 02:58 UT before the M8.7 flare expanded wider and 
higher than the PFSS model at the same time. A quantitative analysis is given below. 
Such a difference is not so obvious for the models at 04:10 UT. As for the magnetic 
connections between different active regions, we find that NOAA 11401 plays a central 
role in this active region group. NOAA 11402 and NOAA 11401 form a long loop arcade. Some 
field lines emanating from NOAA 11401 fall into NOAA 11405. There are also some 
field lines emanating from the negative polarity of NOAA 11401 and 11407 open to the 
interplanetary space.

\begin{figure}
\includegraphics[width=0.7\textwidth]{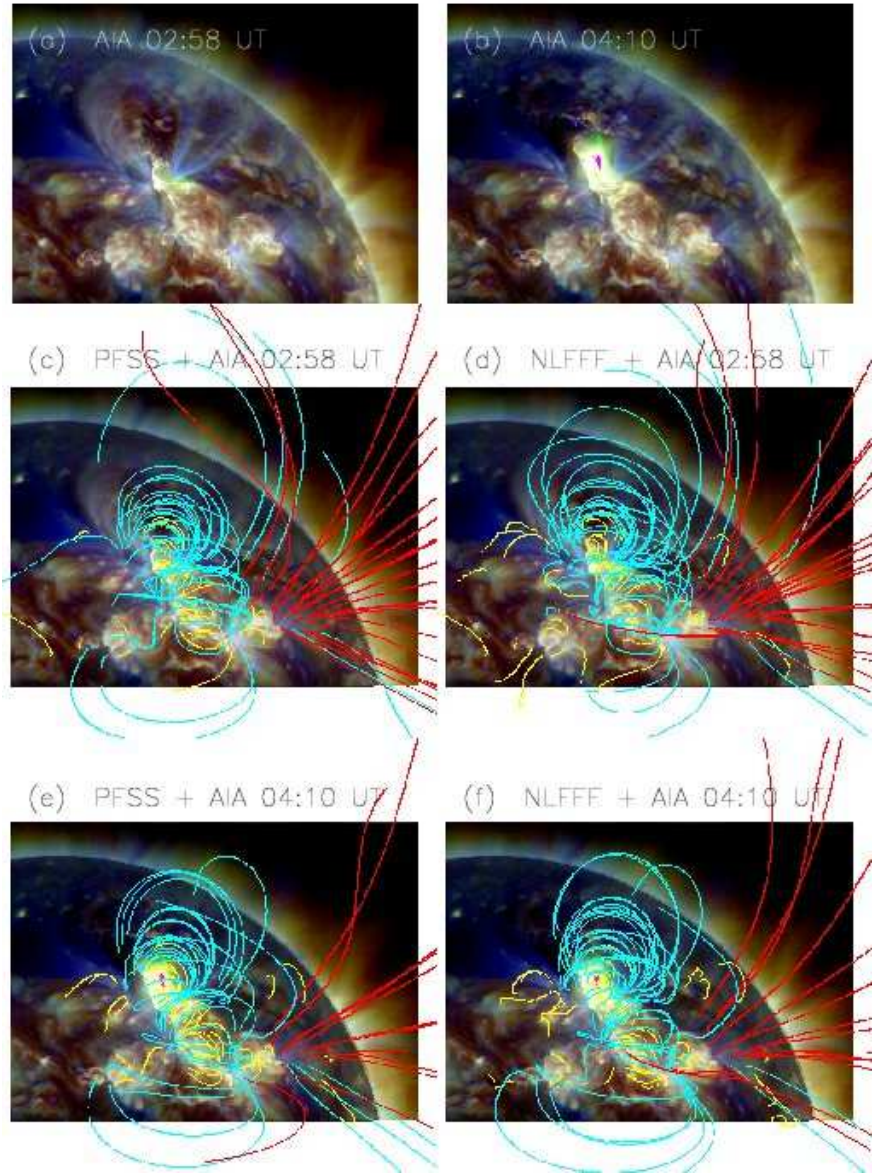}
\caption{(a) The EUV image at 211 \AA \ , 193 \AA \ , and 171 \AA \ bands (corresponding to red, green, and blue
channels, respectively) observed by \textit{SDO}/AIA at 02:58 UT on 2012 January 23. (b) The EUV image
observed at 04:10 UT. (c) Field lines of the PFSS model at 02:58 UT overlaid on the EUV image. The color
code is the same as that in Figure~\ref{fig:nlfff}.
(d) Field lines of the NLFFF model at 02:58 UT overlaid on the EUV image. (e) Field lines of the PFSS model at 
04:10 UT overlaid on the EUV image. (d) Field lines of the NLFFF model at 04:10 UT overlaid on the EUV image.
} \label{fig:aia}
\end{figure}

In order to evaluate whether the NLFFF models are consistent with the observed coronal loops,
and to gauge how much better the NLFFF models are when compared with the associated PFSS 
models, we plot the magnetic
field lines over the \textit{SDO}/AIA EUV images as shown in Figure~\ref{fig:aia}. The
center of the view point is located at $(L_0, B_0)$, where $L_0 = 0.0^\circ$ and $B_0
\approx -5.3^\circ$ are the longitude and latitude of the center of the solar disk,
respectively. The composite EUV image from three AIA channels (211 \AA , 193 \AA , and 171 \AA ,
whose characteristic $\log(T)$ correspond to 6.3, 6.1, and 5.8, respectively)
at 02:58 UT in Figure~\ref{fig:aia}(a) shows that the magnetic loops in active region 
NOAA 11402 are highly nonpotential, since the distance between the footpoints are
less than the height of the loops, and the core region is much brighter than the 
surroundings, which is an indication of large electric current. The PFSS field lines in
Figure~\ref{fig:aia}(c) have an obvious deviation from the observed EUV loops, since the
projection of the field lines leans towards the west, while the loops towards the east. 
The spatial correspondence between the overall shape of the NLFFF field lines and the EUV loops 
is much improved as shown in Figure~\ref{fig:aia}(d). Therefore, the qualitative comparison
between the model magnetic field lines and the observed EUV loops indicates that the
NLFFF model provides a more consistent field for this active region. We recognize that 
the qualitative nature of this comparison is not ideal, however it remains the best option
in the absence of a more reliable quantitative alternative.

After the M8.7 flare, the high EUV loops became invisible,
while the post-flare loop system was more conspicuous as shown in Figure~\ref{fig:aia}(b).
The difference between the magnetic field lines of the PFSS and NLFFF models at 04:10 UT
is not so obvious as shown in Figures~\ref{fig:aia}(e) and (f), which is consistent with
the fact that the electric current has been ejected along with the flare and filament eruption.
The NLFFF at 04:10 UT contains less free magnetic energy than that at 02:58 UT as listed in 
Table~\ref{tbl:hmi}, which provides additional evidence that the overall magnetic field
configuration relaxed to a more potential state after the flare.

\begin{figure}
\includegraphics[width=1.0\textwidth]{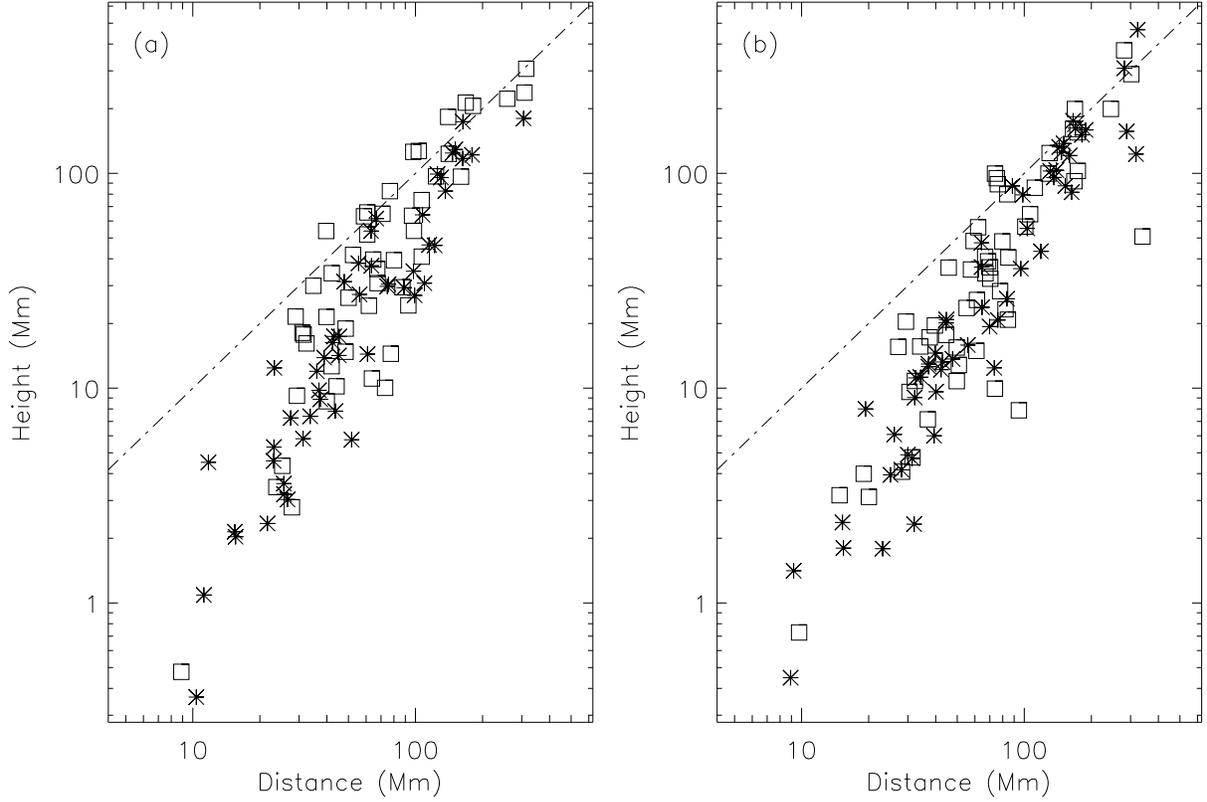}
\caption{The distance of the two footpoints versus the maximum height of a magnetic field line. The asterisks
are for the case of the PFSS field lines, and the squares for the NLFFF field lines. The dash-dotted line
marks the locations where distances equal to heights. (a)
The magnetic field lines are computed with the data at 02:58 UT, and integrated from the same footpoints 
as shown in Figures~\ref{fig:nlfff}(a), \ref{fig:nlfff}(b), \ref{fig:aia}(c) and \ref{fig:aia}(d), but in the 
region $\theta \in [50.0^\circ,70.0^\circ]$ and $\phi \in [10.0^\circ,30.0^\circ]$. (b) The magnetic field lines are 
computed with the data at 04:10 UT, and integrated from the same footpoints as shown 
in Figures~\ref{fig:nlfff}(c), \ref{fig:nlfff}(d), \ref{fig:aia}(e) and \ref{fig:aia}(f), but in the 
region $\theta \in [50.0^\circ,70.0^\circ]$ and $\phi \in [10.5^\circ,30.5^\circ]$. 
} \label{fig:dvsh}
\end{figure}

In order to give a quantitative comparison between the PFSS and NLFFF field lines, we measure
the distance of field line footpoints and the maximum height of a field line. The ratio between
them tells roughly the shape of a field line. The sample field lines are the same as those in 
Figures~\ref{fig:nlfff} and \ref{fig:aia}, but emanating from the active region NOAA 11402.
The distance is measured along the great circle on the solar surface. Figure~\ref{fig:dvsh}
shows the footpoint distance versus the maximum height of the field line for both the PFSS 
and NLFFF models before and after the M8.7 flare. For the field lines higher than 50 Mm 
above the solar surface at 02:58 UT, there are quite some NLFFF field lines whose maximum heights 
are greater than their footpoint distances, but there are very few PFSS field lines having 
this property. Such a behavior becomes weaker for the field lines at 04:10 UT. Therefore, we have a 
quantitative measurement that some NLFFF field lines in the flare productive region expanded
higher than the PFSS field lines before the flare.

\section{Summary and Discussions} \label{sec:conclusion}

We test a NLFFF optimization code in the spherical geometry with the Low and Lou analytical 
solution. It is found that the analytical solution can be recovered to a high level of 
accuracy if vector data on all boundaries are provided. For a more realistic simulation, we only 
provide the vector magnetic field on the bottom. The results show that a weighting function for 
the lateral and top boundaries and a pre-processing for the noisy bottom boundary are necessary 
for a reliable NLFFF extrapolation. Analytical tests also show that the NLFFF code in the 
spherical geometry performs better than that in the Cartesian one when the field of view of 
the bottom boundary is large, say, $20^\circ \times 20^\circ$.

We apply the NLFFF model to an active region observed by \textit{SDO}/HMI both before and after 
an M8.7 flare. The initial condition for the NLFFF optimization method in the spherical geometry 
is computed by the PFSS model. PFSS uses a global radial magnetic field as its boundary 
condition, which can be derived by different methods, such as the synoptic frame or the 
flux-dispersal model. For each observation time, we compare the two results computed with different 
initial conditions derived by the PFSS model using the synoptic frame and the flux-dispersal model, 
respectively. The results show that NLFFF extrapolations using the flux-dispersal model as the 
boundary condition (to compute their initial conditions) have slightly lower, therefore better, force-free 
and divergence-free metrics, and contain larger free magnetic energy. However, if we consider the errors 
as listed in Table~\ref{tbl:hmi} and discussed in Section~\ref{sec:result}, the metrics are 
marginally comparable.

We compare the extrapolated magnetic field lines both from the NLFFF and PFSS models with the
EUV observations by \textit{SDO}/AIA. The comparison shows that NLFFF performs better than PFSS
not only for the core field of the flare productive region but also for the large EUV loops 
higher than 50 Mm. On the one hand, NLFFF bears larger magnetic energy than PFSS. On the other
hand, the EUV loops in active region NOAA 11042, where the M8.7 flare occurred, have non-potential
shapes. NLFFF field lines reproduce these EUV loops better than PFSS before the flare. In this 
active region group, we have both closed and open field lines, both bipolar and quadrupolar 
magnetic configurations. Thus, it is a good example for us to further study the flare itself
and its associated filament eruption and CME. After the flare, the difference between NLFFF
and PFSS becomes smaller than that before the flare.

Compared to previous studies on NLFFF extrapolations in spherical geometry (e.g., 
\citealt{2007Wiegelmann,2009Tadesse,2011Tadesse}), the following additional experiments 
are conducted in this study. First, we quantitatively compare the extrapolation results 
derived by the NLFFF codes both in the spherical and Cartesian coordinates using the Low 
and Lou analytical solution. These comparisons show that geometrical factors need to be
taken into account when the field of view of the bottom boundary is large. Next, we have 
initialized the NLFFF calculations with global PFSS extrapolations, in light of the fact 
that the active regions are rarely isolated and the global PFSS extrapolations provide 
some information about the connectivity of field lines between the active region of interest 
and points outside of the computational domains considered here. Additionally, we test the 
sensitivity of the resulting NLFFF models on the surface fields, and perform comparisons 
between NLFFF extrapolations based on different surface fields (the synoptic frame and 
flux-dispersal model). Finally, we apply the NLFFF extrapolation in spherical geometry to 
\textit{SDO}/HMI observations and compare the magnetic field lines with EUV coronal loops 
observed by \textit{SDO}/AIA. We conclude that the NLFFF model performs better than the PFSS 
model, especially for highly non-potential active regions before eruptive activities.

Due to the limitation of computation resources, we adopt a spatial resolution of $0.5^\circ$
(note that $0.5^\circ$ is the longitude or latitude difference) per grid point for this study. 
Since we focus on the overall property of the core field region, such as the magnetic energy
and the large scale magnetic loops, such a resolution is adequate. However, the \textit{SDO}/HMI 
has a much higher resolution, which is approximately equivalent to $0.03^\circ$ per pixel. In
future research, if one needs to explore the small-scale structure, it would be better to 
interpolate the initial condition to a higher resolution, combine it with the \textit{SDO}/HMI 
vector magnetic fields, and then perform the NLFFF extrapolation to increase the spatial resolution. 

Recently, \citet{2010Wiegelmann} added a new term in the optimization method to 
deal with measurement errors and missing data on the bottom boundary. \citet{2012Wiegelmann} 
has tested this optimization method in the Cartesian geometry with the \textit{SDO} observations, 
and \citet{2012Tadesse} applied the spherical version to observations from the Synoptic Optical 
Long-term Investigations of the Sun (SOLIS). The new optimization method has the advantage
to deal with measurement errors and minimize the force-free and divergence-free terms, while
it takes much longer time for computation than the traditional optimization method. Although
the NLFFF extrapolation techniques have been improved in various ways, it is 
still a challenge to validate all these magnetic field models with coronal observations.

\acknowledgments

The authors thank the anonymous referee for helpful comments.
Data are courtesy of \textit{SDO} and the HMI and AIA science teams. 
Y. Guo thanks J. W. Zhao very much for the invitation to Stanford University. 
Y. Guo and M. D. Ding were supported by NKBRSF under grant 2011CB811402, and 
by NSFC under grants 10878002, 10933003, and 11203014. T. Wiegelmann was 
supported by DLR grant 50 OC 0904.



\begin{table}
\caption{Force-free, divergence-free, vector correlation, Cauchy-Schwartz, normalized vector error, and mean vector error metrics and normalized magnetic energy for the Low and Lou solution, potential field source surface model, and nonlinear force-free field models.}\label{tbl:lowlou}
\begin{tabular}{l l l l l l l l}
\\ \hline \hline
 & $L_\mathrm{f} {}^2$ & $L_\mathrm{d} {}^2$ & $C_\mathrm{vec} {}^2$ & $C_\mathrm{CS} {}^2$ & $1-E_\mathrm{N} {}^2$ & $1-E_\mathrm{M} {}^2$ & $\epsilon {}^2$ \\
\cline{2-3}
Model ${}^1$ & \multicolumn{2}{l}{(10$^{-2}$ G$^2$ Mm$^{-2}$)} & & & & & \\
\hline
Low \& Lou (1990)         & 0.03 & 0.01 & 1    & 1    & 1    & 1    & 1    \\
PFSS ($l_\mathrm{max}=7$) & 0.01 & 0.01 & 0.85 & 0.83 & 0.51 & 0.45 & 0.82 \\
Case 1                    & 0.00 & 0.00 & 1.00 & 1.00 & 0.99 & 0.99 & 1.01 \\
Case 2                    & 0.34 & 0.11 & 0.98 & 0.90 & 0.78 & 0.61 & 0.92 \\
Case 3                    & 0.01 & 0.00 & 0.99 & 0.95 & 0.86 & 0.72 & 1.01 \\
Noisy model 1             & 0.24 & 0.16 & 0.99 & 0.94 & 0.81 & 0.68 & 0.97 \\
Preprocessed 1            & 0.04 & 0.02 & 0.99 & 0.95 & 0.85 & 0.72 & 1.00 \\
Noisy model 2             & 0.91 & 0.61 & 0.97 & 0.90 & 0.72 & 0.59 & 0.99 \\
Preprocessed 2            & 0.11 & 0.06 & 0.99 & 0.94 & 0.84 & 0.71 & 1.00 \\
\hline
\multicolumn{8}{p{15cm}}{\textbf{Notes.}} \\
\multicolumn{8}{p{15cm}}{${}^1$ See Section~\ref{sec:test} for the definitions of the models.} \\
\multicolumn{8}{p{15cm}}{${}^2$ Refer to equations (\ref{eqn:forc})--(\ref{eqn:ener}) in Section~\ref{sec:test} for the definitions of the metrics. 
All the metrics are computed in the computation domain of $r \in [1.0 R_\sun,2.5 R_\sun]$, 
$\theta \in [9.0^\circ,171.0^\circ]$, and $\phi \in [90.0^\circ,270.0^\circ]$.}
\end{tabular}
\end{table}

\begin{table}
\caption{Force-free, divergence-free, vector correlation, Cauchy-Schwartz, normalized vector error, and mean vector error metrics and normalized magnetic energy for the nonlinear force-free field models in Cartesian and Spherical Geometries.}\label{tbl:cartesian}
\begin{tabular}{l l l l l l l l}
\\ \hline \hline
 & $L_\mathrm{f} {}^2$ & $L_\mathrm{d} {}^2$ & $C_\mathrm{vec} {}^3$ & $C_\mathrm{CS} {}^3$ & $1-E_\mathrm{N} {}^3$ & $1-E_\mathrm{M} {}^3$ & $\epsilon {}^3$ \\
\cline{2-3}
Model ${}^1$ & \multicolumn{2}{l}{(10$^{-2}$ G$^2$ Mm$^{-2}$)} & & & & & \\
\hline
Cartesian ($\theta_l = 20^\circ$)   & 96.92 & 57.91 & 0.80 & 0.65 & 0.42 & 0.33 & 0.68 \\
Spherical ($\theta_l = 20^\circ$)   & 0.36 & 0.21 & 0.95 & 0.89 & 0.71 & 0.63 & 1.06 \\
Cartesian ($\theta_l = 40^\circ$)   & 21.98 & 18.02 & 0.89 & 0.80 & 0.50 & 0.40 & 0.98 \\
Spherical ($\theta_l = 40^\circ$)   & 0.16 & 0.12 & 0.98 & 0.94 & 0.82 & 0.72 & 0.99 \\
Cartesian ($\theta_l = 60^\circ$)   & 6.62 & 5.80 & 0.90 & 0.80 & 0.42 & 0.37 & 1.64 \\
Spherical ($\theta_l = 60^\circ$)   & 0.09 & 0.07 & 0.98 & 0.92 & 0.79 & 0.65 & 0.93 \\
\hline
\multicolumn{8}{p{15cm}}{\textbf{Notes.}} \\
\multicolumn{8}{p{15cm}}{${}^1$ See Section~\ref{sec:comparison} for the definitions of the models.} \\
\multicolumn{8}{p{15cm}}{${}^2$ Refer to equations (\ref{eqn:forc}) and (\ref{eqn:dive}) in Section~\ref{sec:test} 
for the definitions of the metrics. $L_\mathrm{f}$ and $L_\mathrm{d}$ are computed in the corresponding 
computation domain. } \\
\multicolumn{8}{p{15cm}}{${}^3$ Refer to equations (\ref{eqn:vect})--(\ref{eqn:ener}) 
in Section~\ref{sec:test} for the definitions of the metrics. 
These metrics are computed in the inner region (excluding a six-grid buffer region) of the corresponding 
cubic box extracted from their computation domain.}
\end{tabular}
\end{table}

\begin{sidewaystable}
\caption{Force-free and divergence-free metrics and magnetic energy for the nonlinear force-free field models and potential field with observed data as the boundary condition.}\label{tbl:hmi}
\begin{tabular}{l l l l l l l l l l l}
\\ \hline \hline
            & Time  &          & $L_\mathrm{f} {}^1$ & $L_\mathrm{d} {}^1$ & $L_\mathrm{f}^\mathrm{pot} {}^1$ & $L_\mathrm{d}^\mathrm{pot} {}^1$ & & $E_\mathrm{m}^\mathrm{nlfff} {}^2$ & $E_\mathrm{m}^\mathrm{pot} {}^2$ & $E_\mathrm{m}^\mathrm{free} {}^2$ \\
\cline{4-7} \cline{9-11}
Date        & (UT)  & Boundary & \multicolumn{4}{l}{(10$^{-2}$ G$^2$ Mm$^{-2}$)} & & \multicolumn{3}{l}{(10$^{33}$ ergs)} \\
\hline
2012 Jan 23 & 02:58 & synoptic frame     & $3.53 \pm 0.06$ & $1.78 \pm 0.01$ & $0.25 \pm 0.00$ & $0.26 \pm 0.00$ & & $1.58 \pm 0.02$ & $1.43 \pm 0.00$ & $0.15 \pm 0.02$ \\
2012 Jan 23 & 02:58 & flux-dispersal$^3$ & $3.41 \pm 0.08$ & $1.76 \pm 0.02$ & $0.25 \pm 0.00$ & $0.26 \pm 0.00$ & & $1.63 \pm 0.03$ & $1.43 \pm 0.00$ & $0.20 \pm 0.03$ \\
2012 Jan 23 & 04:10 & synoptic frame     & $4.76 \pm 0.10$ & $2.20 \pm 0.02$ & $0.32 \pm 0.00$ & $0.33 \pm 0.00$ & & $1.67 \pm 0.01$ & $1.57 \pm 0.00$ & $0.10 \pm 0.01$ \\
2012 Jan 23 & 04:10 & flux-dispersal$^3$ & $4.65 \pm 0.13$ & $2.18 \pm 0.03$ & $0.32 \pm 0.00$ & $0.33 \pm 0.00$ & & $1.69 \pm 0.01$ & $1.57 \pm 0.00$ & $0.12 \pm 0.01$ \\
\hline
\multicolumn{11}{p{15cm}}{\textbf{Notes.}} \\
\multicolumn{11}{p{15cm}}{${}^1$~$L_\mathrm{f}$, $L_\mathrm{d}$, $L_\mathrm{f}^\mathrm{pot}$, 
and $L_\mathrm{d}^\mathrm{pot}$ are the force-free, divergence-free measures for the nonlinear force-free field
and the potential field, respectively. These metrics are computed in the computation domain
including the buffer region. } \\
\multicolumn{11}{p{15cm}}{${}^2$~$E_\mathrm{m}^\mathrm{nlfff}$, $E_\mathrm{m}^\mathrm{pot}$, 
and $E_\mathrm{m}^\mathrm{free}$ are the nonlinear force-free field, potential field, and the free magnetic energy,
respectively, where $E_\mathrm{m}^\mathrm{free} = E_\mathrm{m}^\mathrm{nlfff} - 
E_\mathrm{m}^\mathrm{pot}$. These metrics are computed in the inner region (excluding 
a six-grid buffer region) of their computation domains.} \\
\multicolumn{11}{p{15cm}}{${}^3$~The observed magnetic field has been inserted to the 
flux-dispersal model.}
\end{tabular}
\end{sidewaystable}

\clearpage

\end{document}